\documentclass[aps,pre,floats,twocolumn,showpacs,superscriptaddress]{revtex4}

\usepackage{graphicx}
\usepackage{epsfig}
\usepackage{color}
\usepackage{amsmath}

\begin{document}

\title{Locating privileged spreaders on an Online Social Network}

\author{Javier Borge-Holthoefer}
\affiliation{Instituto de Biocomputaci\'on y F\'\i sica de Sistemas 
Complejos (BIFI), Universidad de Zaragoza, Mariano Esquillor s/n, 50018 Zaragoza, Spain}

\author{Alejandro Rivero}
\affiliation{Instituto de Biocomputaci\'on y F\'\i sica de Sistemas 
Complejos (BIFI), Universidad de Zaragoza, Mariano Esquillor s/n, 50018 Zaragoza, Spain}

\author{Yamir Moreno}
\affiliation{Instituto de Biocomputaci\'on y F\'\i sica de Sistemas 
Complejos (BIFI), Universidad de Zaragoza, Mariano Esquillor s/n, 50018 Zaragoza, Spain}
\affiliation{Departamento de F\'{\i}sica Te\'orica, Universidad de Zaragoza, 50009 Zaragoza, Spain}

\date{\today}

\begin{abstract}
Social media have provided plentiful evidence of their capacity for information diffusion. Fads and rumors, but also social unrest and riots travel fast and affect large fractions of the population participating in online social networks (OSNs). This has spurred much research regarding the mechanisms that underlie social contagion, and also who (if any) can unleash system-wide information dissemination. Access to real data, both regarding topology --the network of friendships-- and dynamics --the actual way in which OSNs users interact--, is crucial to decipher how the former facilitates the latter's success, understood as efficiency in information spreading. With the quantitative analysis that stems from complex network theory, we discuss who (and why) has privileged spreading capabilities when it comes to information diffusion. This is done considering the evolution of an episode of political protest which took place in Spain, spanning one month in 2011.
\end{abstract}

\pacs{89.20.Hh,89.65.-s,89.75.Fb,89.75.Hc}

\maketitle

\section{Introduction}
The question about how a piece of information (a virus, a rumor, an opinion, etc.,) is globally spread over a network, and which ingredients are necessary to achieve such a success, has motivated much research recently. The reason behind this interest is that identifying key aspects of spreading phenomena facilitates the prevention (e.g., minimizing the impact of a disease) or the optimization (e.g. the enhancement of viral marketing) of diffusion processes that can reach system wide scales. In the context of political protest or social movements, information diffusion plays a key role to coordinate action and to keep adherents informed and motivated \cite{gonzalez2011}. Understanding the dynamics of such diffusion is important to locate who has the capability to transform the emission of a single message into a global information cascade, affecting the whole system. These are the so-called ``privileged or influential spreaders''. Beyond purely sociological aspects, some valuable lessons might be extracted from the study of this problem. For instance, current viral marketing techniques (which capitalizes on online social networks) could be improved by encouraging customers to share product information with their acquaintances. Since people tend to pay more attention to friends than to advertisers, targeting privileged spreaders at the right time may enhance the efficiency of a given campaign. 

The prominence (importance, popularity, authority) of a node has however many facets. From a static point of view, an authority may be characterized by the number of connections it holds, or the place it occupies in a network. This is the idea put forward in \cite{gupte2011finding}, where the authors seek the design of efficient algorithms to detect particular (sub)graph structures: hierarchies and tree-like structures. Turning to dynamics, a node may become popular because of the attention it receives in short intervals of time \cite{ratkiewicz2010characterizing} --but that is a rather volatile way of being important, because it depends on activity patterns that change in the scale of hours or even minutes. A more lasting concept of influence comprises both a topological --enduring-- ingredient and the dynamics it supports; this is the case of Centola's ``reinforcing signals'' \cite{centola10science} or the $k$-core \cite{kitsak2010identification}, which we follow here. 

In this paper, we approach the problem of influential spreaders taking into consideration data from the Spanish ``15M movement'' \cite{borge2011structural}. This pacific civil movement is an example of the social mobilizations --from the ``Arab spring''  to the ``Occupy Wall-Street'' movement -- that have characterized 2011. Although whether OSNs have been fundamental instruments for the successful organization and evolution of political movements is not firmly established, it is increasingly evident \cite{borge2011structural} that at least they have been nurtured mainly in OSNs (Facebook, Twitter, etc.) before reaching classic mass media. Data from these grassroots movements --but also from less conflictive phenomena in the Web 2.0-- provide a unique opportunity to observe system-wide information cascades. In particular, paying attention to the network structure allows for the characterization of which users have outstanding roles for the success of cascades of information. Our results complement some previous findings regarding dynamical influence both at the theoretical \cite{kitsak2010identification} and the empirical \cite{gonzalez2011} levels. Besides, our analysis of activity cascades reveals distinctive traits in different phases of the protests, which provides important hints for future modeling efforts.

\section{Data: a networked view of the 15M movement}
\label{sec:data}
The ``15M movement'' is a still ongoing civic initiative with no party or union affiliation that emerged as a reaction to perceived political alienation and to demand better channels for democratic representation. The first mass demonstration, held on Sunday May 15 ($D$ from now on), was conceived as a protest against the management of the economy in the aftermath of the financial crisis. After the demonstrations on day $D$, hundreds of participants decided to continue the protests camping in the main squares of several cities (Puerta del Sol in Madrid, Pla\c ca de Catalunya in Barcelona) until May 22, the following Sunday and the date for regional and local elections.

From a dynamical point of view, the data used in this study are a set of messages (tweets) that were publicly exchanged through {\em www. twitter.com}. The whole time-stamped data collected comprises a period of one month (between April 25th, 2011 at 00:03:26 and May 26th, 2011 at 23:59:55) and it was archived by {\em Cierzo Development Ltd.}, a start-up company. To filter out the whole sample and choose only those messages related to the protests, 70 keywords ({\em hashtags}) were selected, those which were systematically used by the adherents to the demonstrations and camps. The final sample consists of 535,192 tweets. On its turn, these tweets were generated by 85,851 unique users (out of a total of 87,569 users of which 1,718 do not show outgoing activity, i.e., they are only receivers). See \cite{web15m} for more details.

Twitter is most frequently used as a broadcasting platform. Users subscribe to what other users say building a ``who-listens-to-whom'' network, i.e., that made up of followers and followings in Twitter. This means that any emitted message from a node will be immediately available to anyone following him, which is of utmost importance to understand the concept of activity cascade in the next sections. Such relationships offer an almost-static view of the relationships between users, the ``follower network'' for short. To build it, data for all the involved users were scrapped directly from {\em www.twitter.com}. The scrap was successful for the 87,569 identified users, for whom we also obtained their official list of followers restricted to those who had some participation in the protests. The resulting structure is a directed network, direction indicates who follows who in the online social platform. In practice, we take this underlying structure as completely static (does not change through time) because its time scale is much slower, i.e., changes occur probably in the scale of weeks and months. In-degree $k_{in}$ expresses the amount of users a node is following; whereas out-degree represents the amount of users who follow a node. This network exhibits a high level of reciprocity: a typical user holds many reciprocal relationships (with other users who the node probably knows personally), plus a few unreciprocated nodes which typically point at hubs.

\begin{figure}[!th]
\begin{center}
\includegraphics[width=\columnwidth,clip=0]{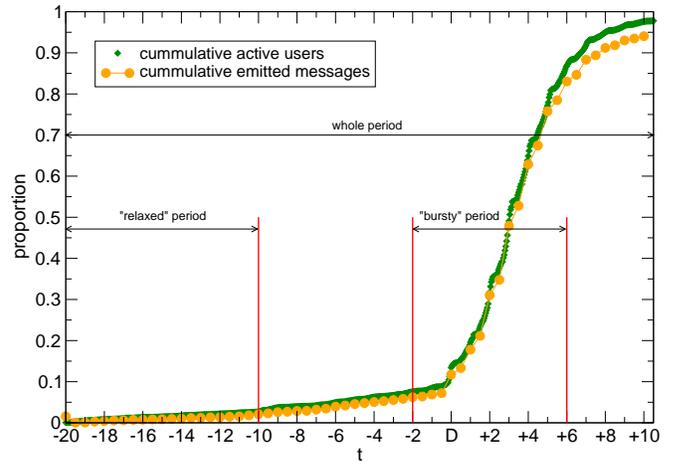}
\caption{(Color online) Temporal evolution of the activity in the online social network. In green, the proportion of nodes that had shown some activity at a certain time $t$. In yellow, the cumulative proportion of emitted messages as a function of time. Note that the two lines evolve in almost the same way. According to this evolution, we have distinguished two sub-periods: one of them characterized as ``slow growth'' due to the low activity level and the other one tagged as ``explosive'' or "bursty" due to the intense information traffic within it.}
\label{growth}
\end{center}
\end{figure}

The main topological features of the follower network fit well in the concept of ``small-world'' \cite{watts98}, i.e., low average shortest path length and high clustering coefficient. Furthermore, both in- and out-degree distribute as a power-law, indicating that connectivity is extremely heterogenous. Thus, the network supporting users' interactions is scale-free with some rare nodes that act as hubs \cite{barabasi99}.

\begin{figure*}[!th]
\centering
\includegraphics[width=0.85\linewidth,clip=0]{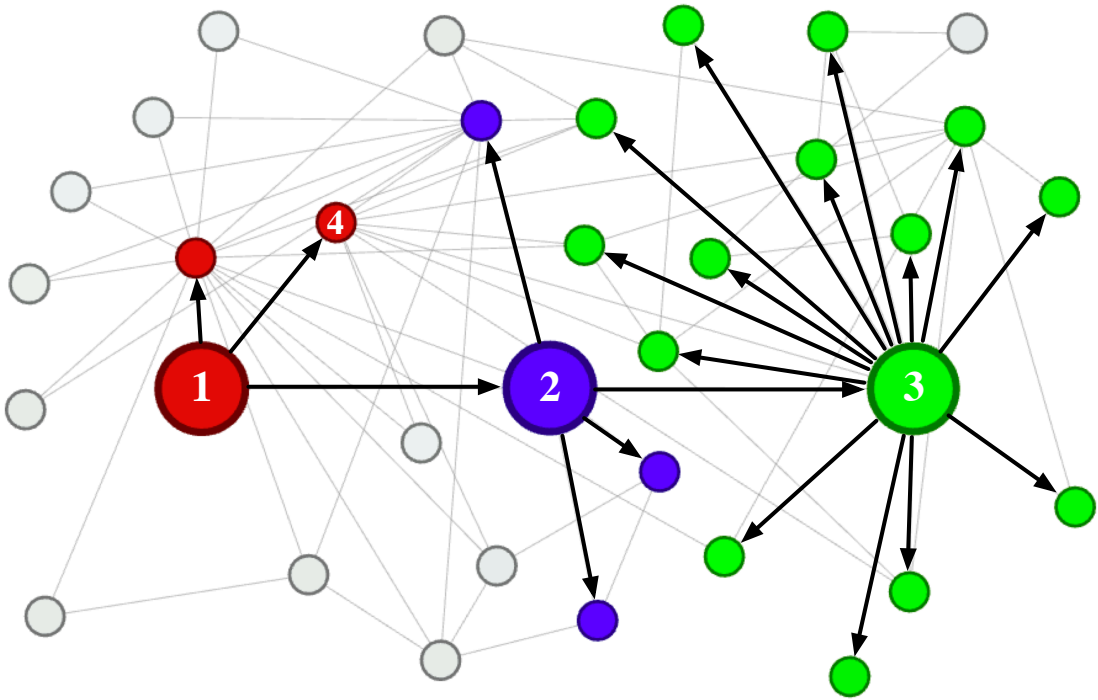}
\caption{(Color online) The figure illustrates the concept of cascade that is used throughout this article. User 1 emits a message at time $t$, and all of his followers automatically receive it. Thus, they are already counted as part of the cascade (small red circles). One of his followers (user 2, big blue node), driven by the previous message, decides himself to participate at time $t+\Delta t$, posting a message himself. A second set of followers are included in the cascade. Finally, a third node (user 3, big green circle) joins in and spreads the cascade further at time $t+2\Delta t$. A node can not be counted twice, note for example that user 4 is also following node 3. Many nodes remain unaffected, because they are not connected to any of the spreaders. The final size of the cascade is $\frac{N_{c}}{N} = \frac{22}{34}$; the success of the cascade largely depends on the capacity to contact a ``leader'' or ``privileged spreader'', i.e., a hub to whom many people listens and who decides to participate. The interesting point, however, is that the number of spreaders needed to attain such success is very low (3), and over 50\% of the cascade is triggered by just one of them.}
\label{example}
\end{figure*}

\section{Methods}
\label{sec:methods}

\subsection{Activity cascades}
An activity cascade --or simply ``cascade'', for short--, starting at a \emph{seed}, occurs whenever a piece of information --or replies to it-- are (more or less unchanged) repeatedly forwarded towards other users. If one of those who ``hear'' the piece of information decides to reply to it, he becomes a \emph{spreader}, otherwise he remains as a mere \emph{listener}. The cascade becomes global if the final number of affected users $N_{c}$ (including the set of spreaders and listeners, plus the seed) is comparable to the size of the whole system $N$. Intuitively, the success of an activity cascade greatly depends on whether spreaders have a large set of followers or not (Figure \ref{example}); remarkably, the seed is not necessarily very well connected. This fact highlights the entanglement between dynamics and the underlying (static) structure.

Note that the previous definition is too general to attain an \emph{operative} notion of cascade. One possibility is to leave time aside, and consider only identical pieces of information traveling across the topology (a {\em retweet}, in the Twitter jargon). This may lead to inconsistencies, such as the fact that a node decides to forward a piece of information long after receiving it (perhaps days or weeks). It is impossible to know whether his action is motivated by the original sender, or by some exogenous reason, i.e., invisible to us. One may, alternatively, take into consideration time, thus considering that, regardless of the exact content of a message, two nodes belong to the same cascade as consecutive spreaders if they are connected (the latter follows the former) and they show activity within a certain (short) time interval, $\Delta t$. The probability that exogenous factors are leading activation is in this way minimized. Also, this concept of cascade is more inclusive, regarding dialogue-like messages (which, we emphasize, are typically produced in short time spans). This scheme exploits the concept of spike train from neuroscience, i.e., a series of discrete action potentials from a neuron taken as a time series. At a larger scale, two brain regions are identified as functionally related if their activation happens in the same time window. Consequently, message chains are reconstructed assuming that activity is contagious if it takes place in short time windows.

\begin{figure*}[!th]
\centering
\includegraphics[width=0.9\linewidth,clip=0]{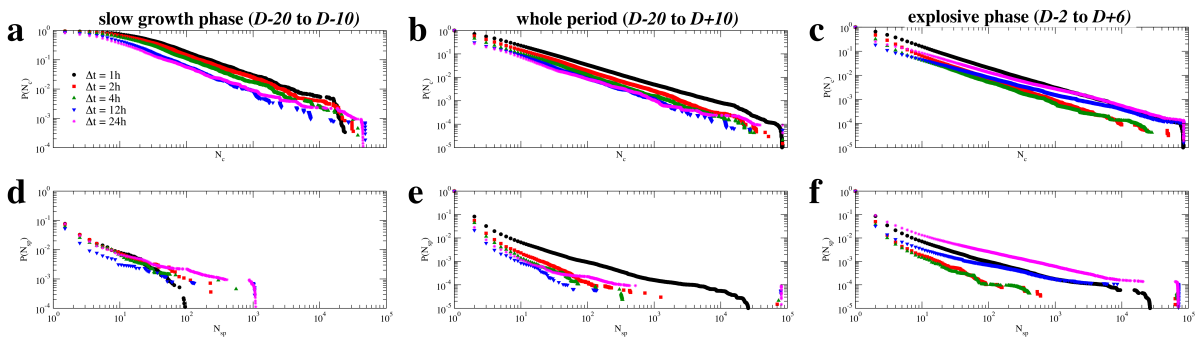}
\caption{(Color online) Upper panels (\emph{a,b,c}): Cascade size probability distributions for the different periods considered. Lower panels (\emph{d,e,f}): Probability distributions of spreaders involved in the cascades for the same periods. The exact periods considered in the analyses are indicated at the top of each panel. See the text for further details.}
\label{fig3}
\end{figure*}

We apply the latter definition to explore the occurrence of information cascades in the data. In practice, we take a seed message posted by $i$ at time $t_{0}$ and mark all of $i$'s followers as listeners. We then check whether any of these listeners showed some activity at time $t_{0}+\Delta t$. This is done recursively until no other follower shows activity, see Figure \ref{example}. In our scheme, a node can only belong to one cascade; this constraint introduces a bias in the measurements, namely, two nodes sharing a follower may show activity at the same time, so their follower may be counted in one or another cascade (with possible important consequences regarding average cascades' size and penetration in time). To minimize this degeneration, we perform calculations for many possible cascade configurations, randomizing the way we process data. We distinguish information cascades (or just cascades, for short) from spreader-cascades. In information cascades we count any affected user (listeners and spreaders), whereas in spreader-cascades only spreaders are taken into account.

We measure cascades and spreader-cascades size distributions for three different scenarios: one in which the information intensity is low (slow growth phase, from $D-20$ to $D-10$), one in which activity is bursty (explosive phase, $D-2$ to $D+6$) and one that considers all available data (which spans a whole month, and includes the two previous scenarios plus the time in-between, $D-20$ to $D+10$). Figure \ref{growth} illustrates these different periods. The green line represents the cumulative proportion of nodes in the network that had shown some activity, i.e., had sent at least one message, measured by the hour. We tag the first 10 days of study as ``slow growth'' because, for that period, the amount of active people grew less than 5\% of the total of users, indicating that recruitment for the protests was slow at that time. The opposite arguments apply in the case of the bursty or ``explosive'' phase: in only 8 days the amount of active users grew from less than 10\% up to over an 80\%. The same can be said about global activity (in terms of the total number of emitted directed messages --the activity network), which shows an almost exact growth pattern. Besides, within the different time periods --slow growth, explosive and total--, different time windows have been set to assess the robustness of our results. Our proposed scheme relies on the contagious effect of activity, thus large time windows, i.e., $\Delta t > 24$ hours, are not considered.

\subsection{$k$-shell decomposition}
The $k$-core decomposition of a network consists of identifying particular subsets of the network, called $k$-cores, each obtained by recursively removing all the vertices of degree less than $k$, where $k = k_{in} + k_{out}$ indicates the total number of in- and out-going links of a node, until all vertices in the remaining graph have degree at least $k$. In the end, each node is assigned a natural number (its coreness), the higher the coreness the closer a node is to the nucleus or core of the network. The main advantage of this centrality measure is, in front of other quantities, its low computational cost that scales as $O(N+E)$, where $N$ is the number of vertices of the graph and $E$ is the number of links it contains \cite{alvarez2008k}. This decomposition has been successfully applied in the analysis of the Internet and the Autonomous Systems structure \cite{alvarez2008k,carmi2007model}. In the following section, we will use the $k$-core decomposition as a means to identify influence in social media. In particular, we discuss which, degree or coreness, is a better predictor of the extent of an information cascade.

\begin{figure*}[!th]
\centering
\includegraphics[width=0.85\linewidth,clip=0]{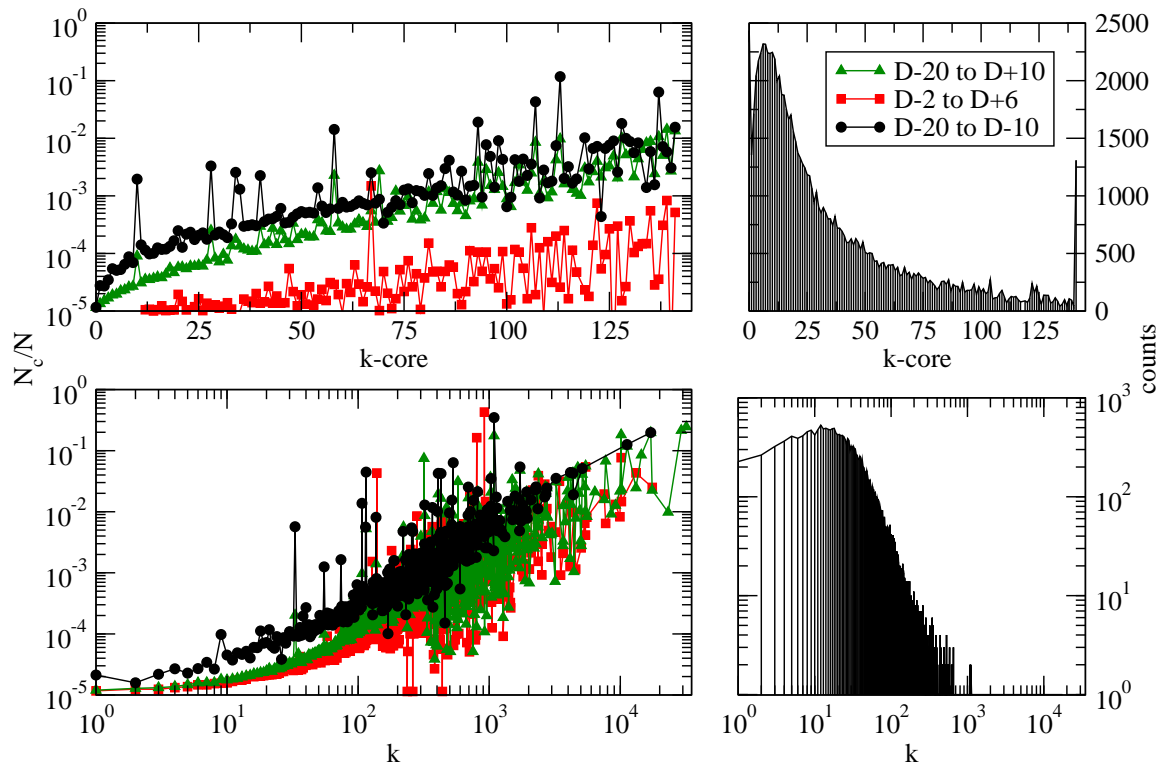}
\caption{(Color online) Left upper panel: average spreading capacity (with respect to the system size) of nodes grouped according to their $k$-core. $\frac{N_{c}}{N}$ grows with coreness, but the explosive period (red squares) evidences a much less clear tendency, with many fluctuations and a lower overall spreading capacity if compared to the slow growth period (black circles). Left lower panel: The same information is showed as a function of the degree. Again, the slow growth period is the best one at predicting the extent of a cascade. Interestingly, average cascades for highest degrees outperform those triggered by highest $k$-core nodes by an order of magnitude. See main text for discussion on this aspect. Right panels show the $k$-core and degree distributions, i.e., how many nodes belong to each class. Note that the highest core contains over 1000 users.}
\label{coredegree}
\end{figure*}

\section{Results}
\label{sec:results}
The upper panels ($a,b,c$) of Figure \ref{fig3} reflect that a cascade of a size $O(N)$ can be reached at any activity level (slow growth, explosive or both). As expected, these large cascades occur rarely as the power-law probability distributions evidence. This result is robust to different temporal windows up to 24h.
In contrast, lower ($d,e,f$) panels show significant differences between periods. Specifically, the distribution of involved spreaders in the different scenarios changes radically from the ``slow growth'' phase (Figure \ref{fig3}d) to the ``explosive'' period (Figure \ref{fig3}f); the distribution that considers the whole period of study just reflects that the bursty period (in which most of the activity takes place) dominates the statistics. The importance of this difference is that one may conclude that, to attain similar results a proportionally much smaller amount of spreaders is needed in the slow growth period. Going to the detail, however, it seems clear (and coherent with the temporal evolution of the protests, Fig. \ref{growth}) that although cascades in the slow period (panel a) affect as much as $N/2$ of the population, the system is in a different dynamical regime than in the explosive one: indeed, distributions suggest that there has been a shift from a subcritical to a supercritical phase.

The previous conclusions raise further questions: is there a way to identify ``privileged spreaders''? Are they placed randomly throughout the network's topology? Or do they occupy key spots in the structure? And, will these influential users be more easily detected in a bursty period (where large cascades occur more often)? In what context will influential spreaders single out? To answer these questions, we capitalize on previous work suggesting that centrality (measured as the $k$-core) enhances the capacity of a node to be key in disease spreading processes \cite{kitsak2010identification}. The authors in \cite{kitsak2010identification} discussed whether the degree of a node (its total number of neighbors, $k$) or its $k$-core (a centrality measure) can better predict the spreading capabilities of such node. Note that the $k$-shell decomposition splits a network in a few levels (over a hundred), while node degrees can range from one or two up to several thousands.

We have explored the same idea, but in relation to activity cascades which are the object of interest here. The upper left panel of Fig.\ \ref{coredegree} shows the spreading capabilities as a function of classes of $k$-cores. Specifically, we take the seed of each particular cascade and save its coreness and the final size of the cascade it triggers. Having done so for each cascade, we can average the success of cascades for a given core number. Remarkably, for every scenario under consideration (slow, explosive, whole), a higher core number yields larger cascades. This result supports the ideas developed in \cite{kitsak2010identification}, but it is at odds with those reported in \cite{borge2012absence}, which shows that the $k$-core of a node is not relevant in rumor dynamics.
Exactly the same conclusion (and even more pronounced) can be drawn when considering degree (lower left panel), which appears to be in contradiction with the mentioned previous evidence \cite{kitsak2010identification}.

At a first sight, our findings seem to point out that if privileged spreaders are to be found, one should simply identify the individuals who are highly connected. However, this procedure might not be the best choice. The right panels in Figure \ref{coredegree} show the $k$-core (upper) and degree (lower) distributions, indicating the number of nodes which are seeds at one time or another, classified in terms of their coreness or degree. Unsurprisingly, many nodes belong to low cores and have low degrees. The interest of these histograms lies however in the tails of the distributions, where one can see that, while there are a few hundred nodes in the high cores (and even over a thousand in the last core), highest degrees account only for a few dozen of nodes. In practice, this means that by looking at the degree of the nodes, we will be able to identify quite a few influential spreaders (the ones that produce the largest cascades). However, the number of such influential individuals are far more than a few. As a matter of fact, high cascading capabilities are distributed over a wider range of cores, which in turn contain a significant number of nodes. Focusing on Fig.\ \ref{coredegree}, note that triggering cascades affecting over $10^{-2}$ of the network's population demands nodes with $k \ge 10^{3}$. Checking the distribution of degrees (right-hand side), it is easy to see that an insignificant amount of nodes display such degree range. In the same line, we may wonder what it takes to trigger cascades affecting over $10^{-2}$ of the network's population, from the $k$-core point of view. In this case, nodes with $k$-core around 125 show such capability. A quick look at the core distribution yields that over 1500 nodes accomplish these conditions, i.e., they belong to the 125th $k$-shell or higher.

We may now distinguish between scenarios in Figure \ref{coredegree}. While any of the analyzed periods shows a growing tendency, i.e., cascades are larger the larger is the considered descriptor, we highlight that it is in the slow growth period (black circles) where the tendency is more clear, i.e., results are less noisy. Between the other two periods, the explosive one (red squares) is distinctly the less robust, in the sense that cascade sizes oscillate very much across $k$-cores, and the final plot shows a smaller slope than the other two. This subtle fact is again of great importance: it means that during ``information storms'' a large cascade can be triggered from anywhere in the network (and, conversely, small cascades may have begun in important nodes). The reason for this is that in periods where bursty activity dominates the system suffers ``information overflow'', the amount of noise flattens the differences between nodes. For instance, in these periods a node from the periphery (low coreness) may balance his unprivileged situation by emitting messages very frequently. This behavior yields a situation in which, from a dynamical point of view, nodes become increasingly indistinguishable. The plot corresponding to the whole period analyzed (green triangles) lies consistently between the other two scenarios, but closer to the relaxed period. This is perfectly coherent, the study spans for 30 days and the explosive period represents only 25\% of it, whereas the relaxed period stands for over 33\%. Furthermore, those days between $D-10$ and $D-2$, and beyond $D+6$, resemble the relaxed period as far as the flow of information is concerned.

\section{Conclusions}
\label{sec:conclusions}
Online social networks are called to play an ever increasing role in shaping many of our habits (be them commercial or cultural) as well as in our position in front of political, economical or social issues not only at a local, country-wide level, but also at the global scale. It is thus of utmost importance to uncover as many aspects as possible about topological and dynamical features of these networks. One particular aspect is whether or not one can identify, in a network of individuals with common interests, those that are influentials to the rest. Our results show that the degree of the nodes seems to be the best topological descriptor to locate such influential individuals. However, there is an important caveat: the number of such privileged seeds is very low as there are quite a few of these highly connected subjects. On the contrary, by ranking the nodes according to their $k$-core index, which can be done at a low computational cost, one can safely locate the (more abundant in number) individuals that are likely to generate large (near to) system-wide cascades. The results here presented also lead to a surprising conclusion: periods characterized by explosive activity are not convenient for the spreading of information throughout the system using influential individuals as seeds. This is because in such periods, the high level of activity --mainly coming from users which are badly located in the network-- introduces noise in the system. Consequently, influential individuals lose their unique status as generators of system wide cascades and therefore their messages are diluted. 

On more general grounds, our analysis of real data remarks the importance of empirical results to validate theoretical contributions. In particular, Fig. \ref{coredegree}, together with the observations in \cite{borge2012absence}, raises some doubts about rumor dynamics as a good proxy to real information diffusion. We hypothesize that such models approach information diffusion phenomena in a too simplistic way, thus failing to comprise relevant mechanisms such as complex activity patterns \cite{barabasi2005origin,fernandez11,vazquez2007impact}. 
Finally, although the underlying topology may be regarded as constant, any modeling effort should also contemplate the time evolution of the dynamics. Indeed, Fig. \ref{fig3} suggests that the system is in a sub-critical phase when activity level is low, and critical or supercritical during the explosive period. This is related to the rate at which users are increasingly being recruited as active agents, i.e. the speed at which listeners become spreaders.


\section*{Acknowledgments}
This work has been partially supported by MICINN through Grants FIS2008-01240 and FIS2011-25167, and by Comunidad de Arag\'on (Spain) through a grant to the group FENOL. 


\end{document}